\documentclass[a4paper]{llncs}

\usepackage{amssymb}
\usepackage{graphicx}

\usepackage{amsmath}

\usepackage{caption}
\usepackage{subfig}

\newcommand{\keywords}[1]{\par\addvspace\baselineskip
\noindent\keywordname\enspace\ignorespaces#1}

\begin{document}

\title{A Multi-signal Variant for the GPU-based Parallelization of
Growing Self-Organizing Networks}

\author{Giacomo Parigi\inst{1} \and Angelo Stramieri\inst{1} \and Danilo
Pau\inst{2} \and Marco Piastra\inst{1}}

\institute{Computer Vision and Multimedia Lab, University of Pavia, Via Ferrata
1 - 27100 Pavia (PV), Italy \\
\email{giacomo.parigi@gmail.com}
\and Advanced System Technology, STMicroelectronics, Via Olivetti 2 - 20864
Agrate Brianza (MB), Italy}

\maketitle

\begin{abstract}

Among the many possible approaches for the parallelization of self-organizing
networks, and in particular of \emph{growing} self-organizing networks, perhaps
the most common one is producing an optimized, parallel implementation of
the standard sequential algorithms reported in the literature. In this paper we
explore an alternative approach, based on a new algorithm variant specifically
designed to match the features of the large-scale, fine-grained parallelism of
GPUs, in which multiple input signals are processed at once. Comparative tests
have been performed, using both parallel and sequential implementations of the
new algorithm variant, in particular for a growing self-organizing network that
reconstructs surfaces from point clouds. The experimental results show that this
approach allows harnessing in a more effective way the intrinsic parallelism
that the self-organizing networks algorithms seem intuitively to suggest,
obtaining better performances even with networks of smaller size.

\end{abstract}

\section{Introduction}
\label{sec:introduction}

From a general point of view a self-organizing network is composed by units that
adapt themselves, through limited and local interactions, to input signals from
some predefined space. In most cases a topology is defined among these units by
a set of binary connections. At first sight, the adaptation process may look
inherently parallel, since each unit follows the same predetermined behavior and
in many cases, as long as two units are sufficiently far away in the network,
they do not interact in any way.

Nonetheless, most of the algorithms in the literature are described as
sequential procedures, in the sense that input signals are submitted one by
one to the network and processed each in a sequential way. This means that,
in most cases, also units will be adapted sequentially, one after the other,
even when they can be considered as mutually independent, i.e. with input
signals that are sufficiently distant in the input space.

In a typical algorithm, each input signal has to be compared to all units in the
network, in order to find the closest one and adapt the latter and its neighbors
to the input signal. For reasons that will be described in detail later on, this
operation is dominant in terms of execution time, and is therefore the obvious
focus for parallel implementation. In this respect, two main methods emerge:
\emph{data partitioning} methods, in which the input signals are partitioned
across parallel tasks, whereby each task involves the entire network and
processes just one input signal; \emph{network partitioning} methods, in which
the units of the network are partitioned across parallel tasks, whereby each
task considers all input signals but only in relation to the units belonging to
its partition. These two approaches are thoroughly examined in
\cite{lawrence1999scalable} for the parallelization of Kohonen's self-organizing
map\cite{kohonen2001self}. In particular, in the former work, a data
partitioning approach is described for the batch version of the algorithm,
and a network partitioning approach for the on-line version of the
algorithm, in both cases for an SP2 parallel computer.

\begin{figure}[t!]
\centering
\includegraphics[width=0.19\textwidth]{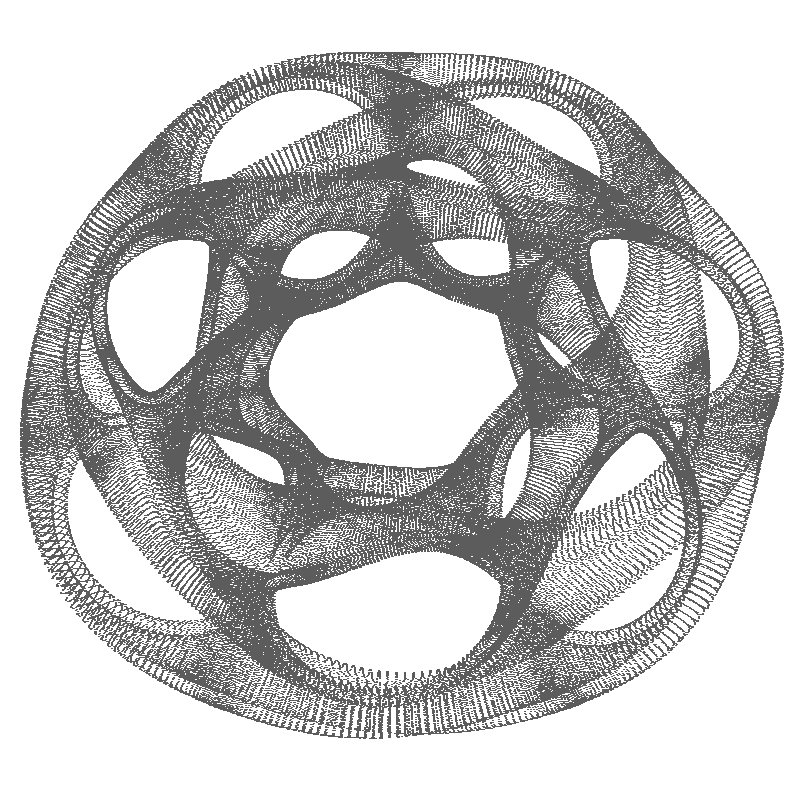}
\includegraphics[width=0.19\textwidth]{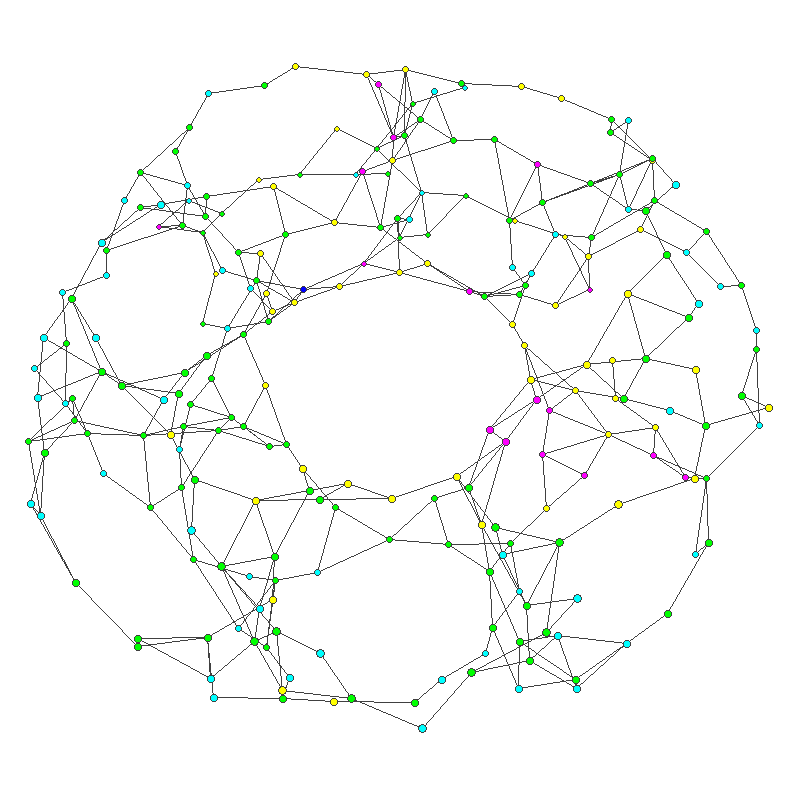}
\includegraphics[width=0.19\textwidth]{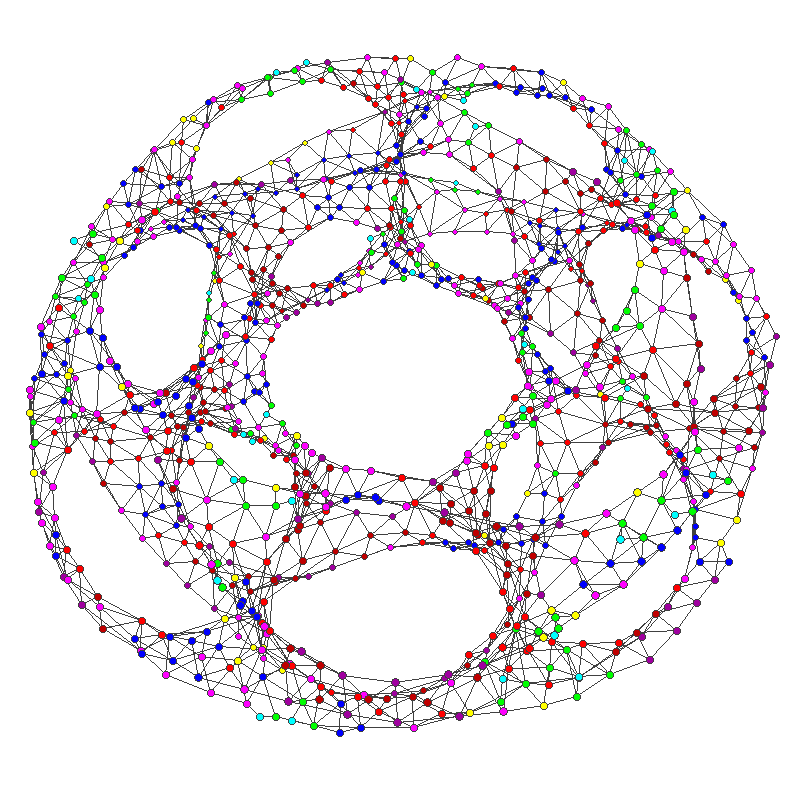}
\includegraphics[width=0.19\textwidth]{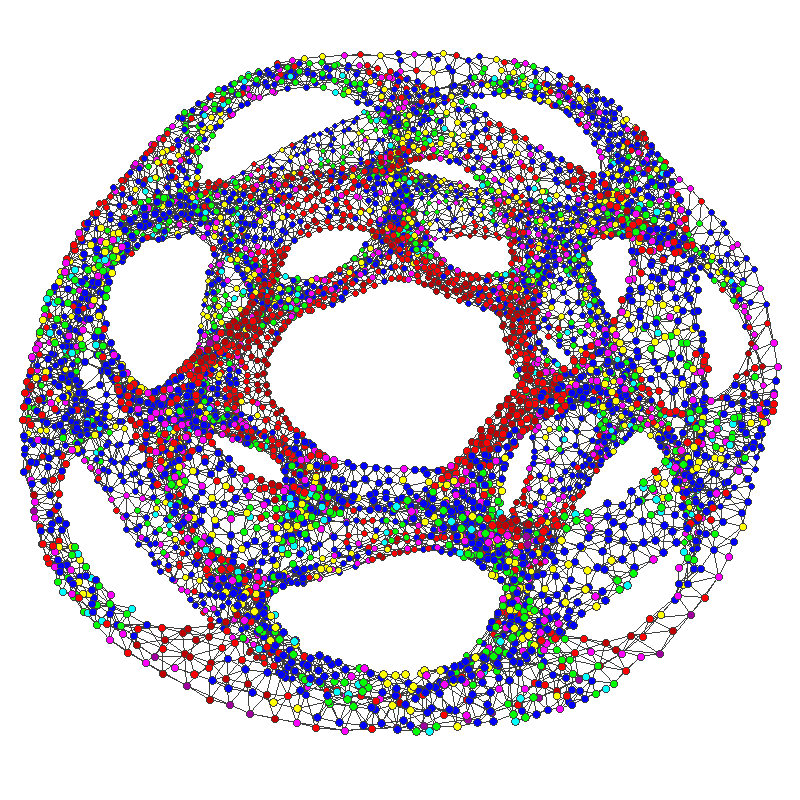}
\includegraphics[width=0.19\textwidth]{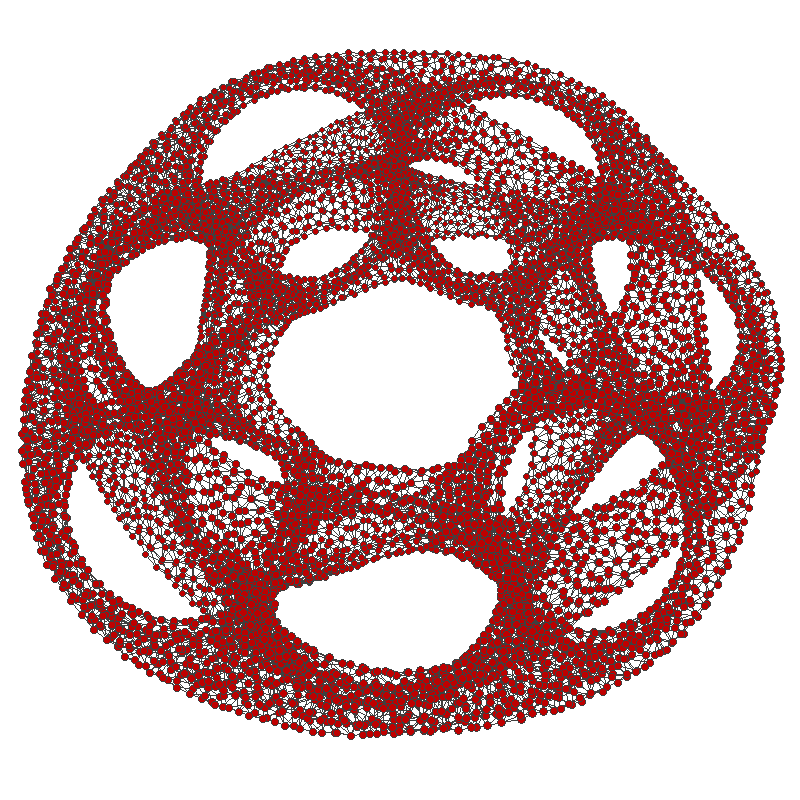}
\caption{The SOAM \cite{piastra2012self} reconstructs a surface from the
point cloud on left. At the end, all units converge to the same stable state.}
\label{fig:soam}
\end{figure}

As a matter of fact, perhaps the most common approach for the parallel
implementation of self-organizing networks described in the literature (see
for instance \cite{orts2012gpgpu}, \cite{garcia2011fast},
\cite{luo2005artificial}, \cite{campbell2005graphics}), is to adapt the
network-partitioning method to the standard, sequential version of the
algorithm.

In an effort to better harness the ``latent parallelism'' of self-organizing
networks, a new algorithm variant for \emph{growing} self-organizing networks is
proposed in this paper. In this \emph{multi-signal} algorithm variant, a number
of signals are submitted to the network and elaborated at once during each
iteration. This variant is explicitly intended for a data-partitioning approach
to parallelization, which, as described in \cite{lawrence1999scalable}, may
offer ``the potential for much greater scalability, since the parallel granularity is
determined by the volume of data, which is potentially very large''. In
particular the new algorithm focuses on \emph{growing} self-organizing networks
and this entails dealing with some further functional aspects, that are not
present in the algorithm for self-organizing maps considered in
\cite{lawrence1999scalable}. This aspect will be described in section
\ref{sec:methods}.

The new multi-signal algorithm has been designed to match the features of the
large-scale, fine-grained parallelism of GPUs (Graphics Processing Units).
Beside its computational capabilities, this hardware has gained a large
popularity due to the lower costs compared to those of more traditional
high-performance computing solutions. For instance, in \cite{owens2007survey},
GPUs have been defined ``probably today's most powerful computational hardware
for the dollar''.

The GPU-based implementation of the multi-signal variant, has shown good
performances in all the tests performed, reaching noticeable speed-ups even for
smaller networks. In addition, the new multi-signal algorithm has shown some
interesting differences w.r.t. the standard single-signal algorithm: in the
tests performed, the multi-signal algorithm always required less input signals
to reach termination than the single-signal counterpart. These aspects will be
further discussed in section \ref{sec:results}.

\section{Methods}
\label{sec:methods}

\subsection{Growing Self-Organizing Networks}
\label{sec:self}

In the discussion that follows, we consider as reference a network in which each
unit is associated to a \emph{reference vector} in the input space, and a
topology is defined by a set of binary connections between the units.
These connections also define the local topology, or \emph{neighborhood}, of
each unit. In a self-organizing network units are progressively adapted during
the learning process. In addition, \emph{growing} self-organizing networks, like
GNG \cite{fritzke1995growing}, GWR \cite{marsland2002self} and SOAM
\cite{piastra2012self} have the following characteristics:

\begin{itemize}
    \item during the learning process the number of units varies, and can
    both grow and shrink;
    \item the topology of connections between units varies as well, since
    connections are both created and destroyed during the learning process.
\end{itemize}

In general, the learning process of a growing self-organizing network can be
described as a basic iteration, which is repeated until some convergence
criterion is met:

\begin{enumerate}
  \item \emph{Sample}\\
  Generate at random one input signal $\boldsymbol{\xi}$ with probability
  $P(\boldsymbol{\xi})$. Usually the support of $P(\boldsymbol{\xi})$ coincide
  with the \emph{region of interest}, i.e. the region of input space to be
  considered.
 
  \item \emph{Find Winners}\\
  Compute the distances $\| \boldsymbol{\xi} - \mathbf{w}_i \|$ between each
  reference vector and the input signal and find the $k$-nearest units. In
  most cases $k=2$, i.e. the nearest (\emph{winner}) and second-nearest 
  units are searched for.
  
  \item \emph{Update the Network}\\
  Create a new connection between the winner and the second-nearest unit, if not
  existing, or reset the existing one\footnote{An \emph{aging} mechanism is also
  applied to connections (see for instance \cite{fritzke1995growing}).}.\\
  Adapt the reference vector of the winner unit and of its topological
  neighbors, with a law of the type:
  \begin{equation}
  \label{eq:adaptation}
  \begin{split}
     \Delta\mathbf{w}_b &= \varepsilon_b \|\boldsymbol{\xi} - \mathbf{w}_b\|,\\
     \Delta\mathbf{w}_i &= \varepsilon_i \eta(i,b) \|\boldsymbol{\xi} - \mathbf{w}_i\|,
  \end{split}
  \end{equation}
  where $\mathbf{w}_b$ is the reference vector of the winner and $\mathbf{w}_i$
  are the reference vectors of the neighboring units. $\varepsilon_b,
  \varepsilon_i, \in \left[0, 1\right]$ are the \emph{learning rates}, with
  $\varepsilon_b \gg \varepsilon_i$. The function $\eta(i,b) \leq 1$ determines
  how neighboring units are adapted. In most cases $\eta(i,b) = 1$ if units $b$
  and $i$ are connected and $0$ otherwise. During this phase, new units can
  be created and old units can be removed, with methods that may vary depending
  on the specific algorithm.
\end{enumerate} 

\begin{figure}[t!]
\centering
\includegraphics[width=0.75\textwidth]{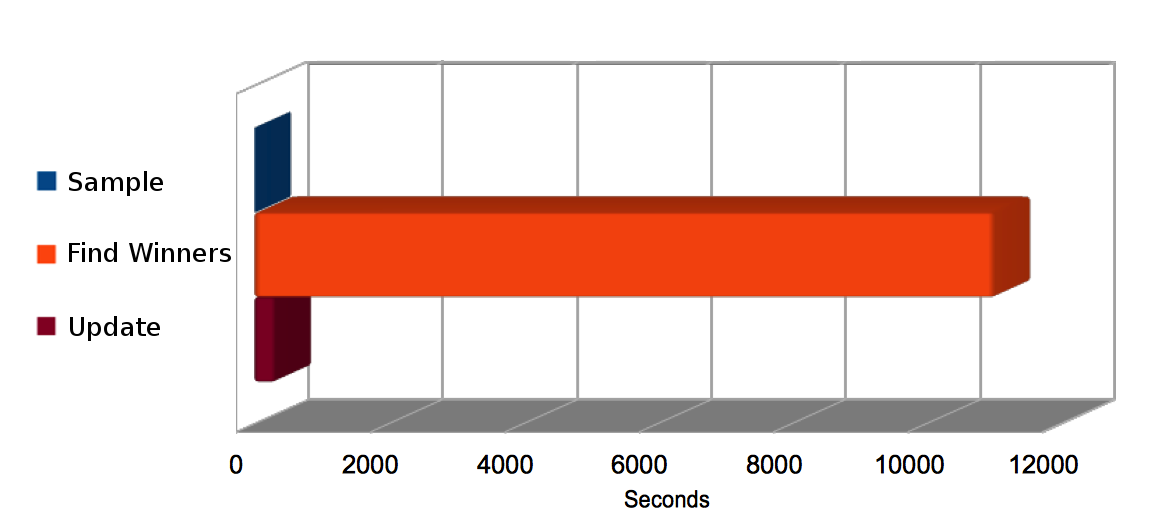}
\caption{Single-phase time to convergence of the SOAM algorithm (average
values on the whole test set).}
\label{fig:phases}
\end{figure}

The three steps above are iterated until some convergence criterion is met:
tipically, in most self-organizing networks, including growing ones, this
criterion is a threshold on the overall quantization error, i.e. the mean
squared distance between input signals and the corresponding winners. For the
purposes of this work we adopted the SOAM algorithm, that has a termination
criterion which does not depend on a threshold. In the SOAM algorithm, in fact,
the learning process terminates when all units have reached a local topology
consistent with that of a surface and therefore the network represents the
triangulation of the surface that has to be reconstructed from input signals
(see Fig.\ref{fig:soam}). The clear termination criterion in the SOAM algorithm
is fundamental for comparing the performances and the different behaviors of the
two variants of the algorithm, i.e. single-signal and multi-signal, and their
implementations.

The methods adopted for adding and removing units during the \emph{Update} phase
greatly vary depending on the algorithm. In GNG, for example, new units are
inserted at regular intervals, in the neighborhood of the unit $i$ that has
accumulated the largest mean squared error. In contrast, in GWR new units are
added whenever the distance between the winner unit and the input signal
$\boldsymbol{\xi}$ is greater than a predefined \emph{insertion threshold}.
The new unit is positioned in proximity of the winner and the network topology
is updated. The SOAM algorithm is similar to the GWR, with the difference that
the insertion threshold may vary during the learning process, in order to
reflect the \emph{local feature size} (LFS) of the surface being reconstructed
(see section \ref{sec:comparison}).

In terms of time complexity, the \emph{Find Winners} phase is dominant in
general. In fact, assuming that the number $k$ of nearest neighbors is constant
and small, the \emph{Find Winners} phase has $\mathcal{O}(N)$ time complexity,
where $N$ is the total number of units. Although the complexity of the
\emph{Update} phase may greatly vary depending on how the function is defined
(see for instance the Neural Gas algorithm \cite{martinetz1994topology}), as a
matter of fact in most growing self-organizing networks, including the SOAM
algorithm, this update is local and limited to the connected neighbors of the
winner, so that the \emph{Update} phase can be assumed to have $\mathcal{O}(1)$
time complexity. Furthermore, in this discussion, we will not consider the
\emph{Sample} phase in detail: sampling methods, in fact, are
application-dependent and not necessarily under the control of the algorithm.

The dominance of the \emph{Find Winners} phase in terms of time complexity is
confirmed by experiments. The graph in Fig.\ref{fig:phases} shows the mean
values obtained from the experiments described in section \ref{sec:results}.
These results are in line with the ones reported in the literature (see for
example \cite{orts2012gpgpu} for a detailed analysis), in that the percentage
of the execution time spent in the \emph{Find Winners} phase remains as low as
50-60\% of the total execution time as long as the network remains small
(i.e. 250-500 units), but grows rapidly to 95\% and more as the network
size increases and more signals are processed.

\subsection{The Multi-signal Variant}
\label{sec:multi}

In the \emph{multi-signal} variant proposed here, at each iteration $m \gg 1$
signals are considered at once. The basic iteration hence becomes:

\begin{enumerate}
  \item \emph{Sample}\\
  Generate at random $m$ input signals $\boldsymbol{\xi}_1, \ldots,
  \boldsymbol{\xi}_m$ according to the probability distribution
  $P(\boldsymbol{\xi})$, as described before.
  
  \item \emph{Find Winners}\\
  For each signal $\boldsymbol{\xi}_j$, compute the distances $\|
  \boldsymbol{\xi}_j - \mathbf{w}_i \|$ between each reference vector and the
  input signal and find the $k$-nearest units.
  
  \item \emph{Update the Network}\\
  For each signal $\boldsymbol{\xi}_j$, perform the update operations specified
  in the previous section.
\end{enumerate}

The first two phases in the above iteration pose no particular problems, since
all the involved operations performed for each signal are mutually independent.
In contrast, during the \emph{Update} phase, the operations performed for
different signals may collide. In particular three kinds of collision can occur:

\begin{figure}[t!]
\centering
\subfloat[][\label{fig:adaptCollision1}]{
\includegraphics[width=0.32\textwidth]{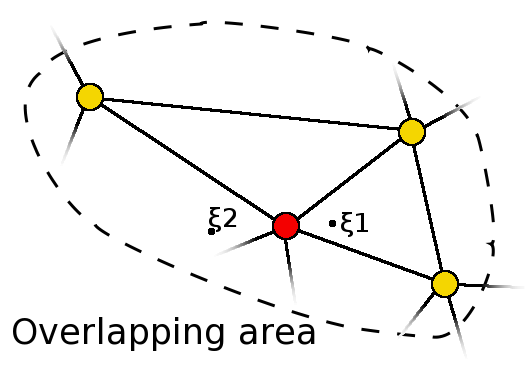}}
\subfloat[][\label{fig:adaptCollision2}]{
\includegraphics[width=0.32\textwidth]{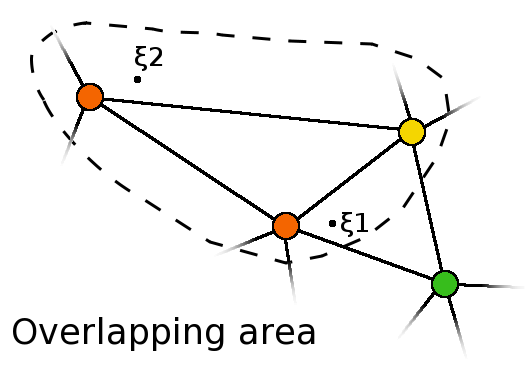}}
\subfloat[][\label{fig:adaptCollision3}]{
\includegraphics[width=0.32\textwidth]{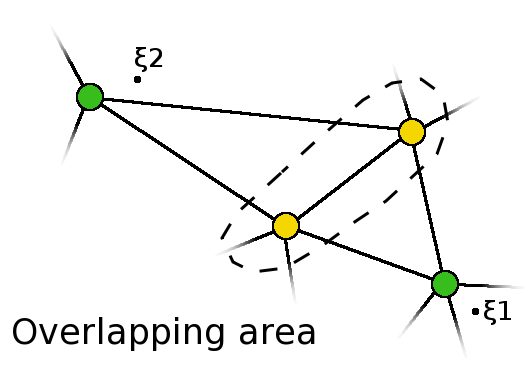}} \caption{
Collision caused by two different input signals $\xi_1$ and $\xi_2$. In (a) the
two signals share the same winner, hence all direct neighbors. In (b) and (c)
the winner is different, but the neighbors are shared. All three cases would
lead to colliding adaptations.}
\label{fig:adaptCollision}
\end{figure}
\begin{description}
  \item [\emph{Adapt position}.] Two or more signals lead to the
  adaptation of the same unit in the network. Collisions of this kind
  are usually not isolated, since the collision can happen both for the winner
  and for the neighboring units, as described in Fig. \ref{fig:adaptCollision}.
%
%
  \item [\emph{Modify neighborhood}.] Two or more signals lead to modifying the
  neighborhood of the same unit. This may be caused by either the
  insertion/removal of units or the creation/removal of edges.
  \item [\emph{Insert edge}.] Two or more signals lead to the
  insertion of the same edge.
\end{description}

In the multi-signal variant presented here, the main concern in choosing the
method for managing the above collisions is maintaining a coherent behavior with
respect to the single-signal algorithm, and allow an unbiased comparison of the
performances. At the same time, the method must be simple enough. The solution
adopted is using an implicit \emph{lock} on the winner unit: no two or more
input signals having the same winner can cause any of the colliding
modifications to be performed, as only the first incoming signal, in a random
order, will produce the corresponding effect, while other signals for the same
winner will just be \emph{discarded}.

Collisions apart, the behavior of the new variant is \emph{different} from the
original, single-signal algorithm. In the experiments described in section
\ref{sec:results}, in fact, the multi-signal variant always required a smaller
number of signals to reach convergence, regardless of the implementation. This
aspect will be discussed in more detail in section \ref{sec:behavior}. 

\begin{figure}[t!]
\centering
\includegraphics[width=0.60\textwidth]{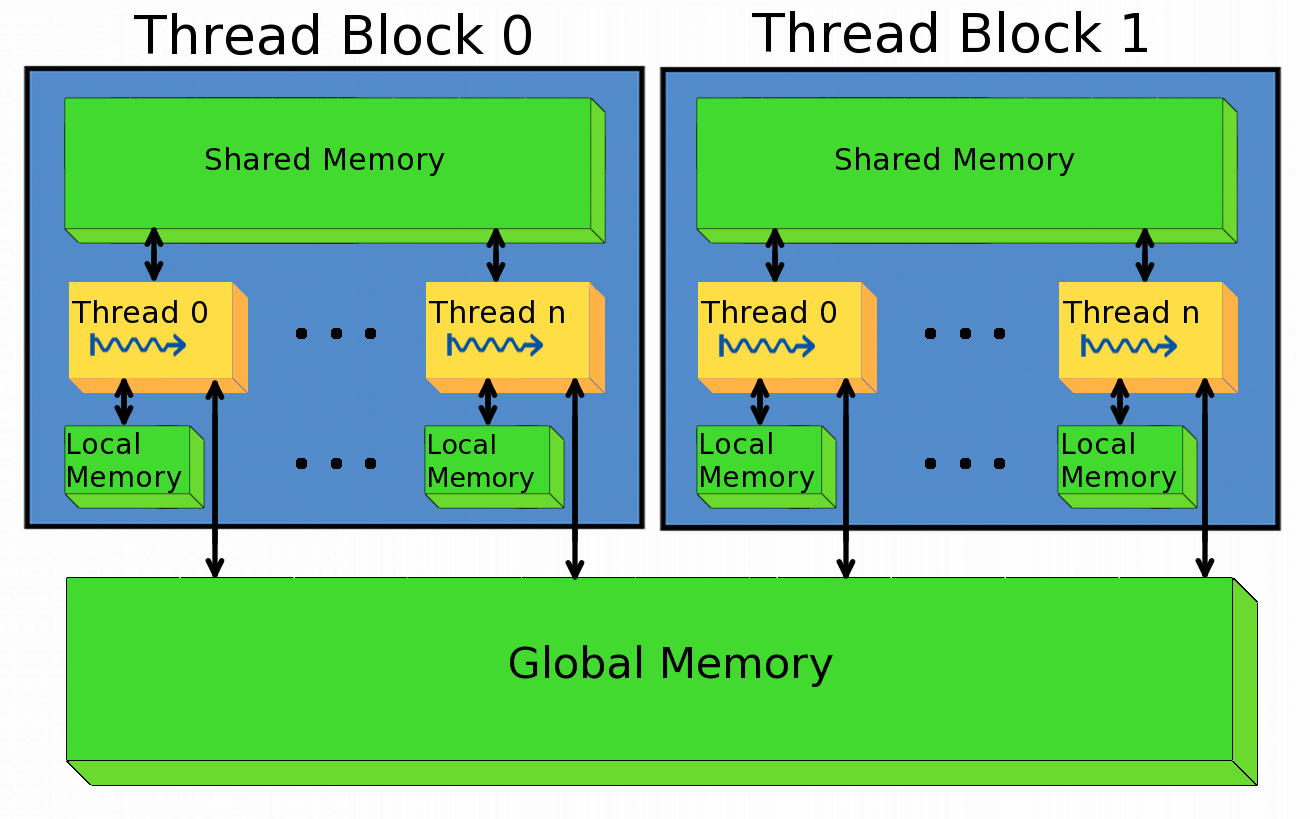}
\caption{Standard GPU memory hierarchy}
\label{fig:hierarchy}
\end{figure}

\subsection{Graphics Processing Units}
\label{sec:gpu}

Graphics Processing Units (GPUs) are specialized and optimized for graphic
applications, and are typically mounted on dedicated boards with private onboard
memories. During these last years, GPUs have evolved into general-purpose
parallel execution machines \cite{owens2008gpu}. Until not many years ago, in
fact, the only available programming interfaces (API) for GPUs were very
specific, forcing the programmer to translate the task into the graphic
primitives provided. Gradually, many general-purpose API for parallel computing
have emerged, which are suitable for GPUs as well. This set of API includes,
for instance, RapidMind \cite{mccool2006data}, PeakStream
\cite{papakipos2007peakstream} or the programming systems owned by NVIDIA and
AMD, respectively CUDA (Compute Unified Device Architecture)
\cite{nvidia2011nvidia} and CTM (Close to Metal) \cite{hensley2007amd}, together
with proposed vendor-independent standards like OpenCL \cite{stone2010opencl}.

Albeit with some differences, all these API adopt the general model of
\emph{stream computing}: data elements are organized in \emph{streams}, which
are ordered sets of data; a set of streams is processed by the same
\emph{kernel}, intended as a set of functions to be computed in parallel, and
produces another set of streams as output. Each kernel is executed on a set of
GPU cores in the form of concurrent \emph{threads}, each executing the same
program on a different stream of data. Within a kernel, threads are grouped into
blocks and each block is executed in sync. In case of branching of the
execution, the block is partitioned in two: all the threads on the first branch
are executed in parallel and then the same is done for all the threads in the
second branch. This general model of parallel execution is often called SIMT
(single-instruction multiple-thread) or SPMD (single-program multiple-data);
compared to the older SIMD, it allows greater flexibility in the flow of
different threads, although at the cost of a certain degree of serialization,
depending on the program. This means that, although independent thread
executions are possible, blocks of coherent threads with limited branching will
make better use of the GPU's hardware.
In typical GPU architectures, onboard and on-chip memories are organized in a
hierarchy (Fig.\ref{fig:hierarchy}): \emph{global} memory, i.e. accessible by
all threads in execution, \emph{shared} memory, i.e. a faster cache memory
dedicated to each single thread block and \emph{local} memory and/or registers, which are
private to each thread.

Another noteworthy feature of modern GPUs is the wide-bandwidth access to
onboard memory, on the order of 10x the memory access bandwidth on typical PC
platforms. To achieve best performances, however, memory accesses by different
threads should be made aligned and coherent, in order to \emph{coalesce} them
into fewer, parallel accesses addressing larger blocks of memory. Incoherent accesses, on
the other hand, must be divided into a larger number of sequential memory
operations. One of the aspects that make GPU programming still quite complex is
that, in most cases, the hierarchy of levels of memory, in particular the
shared memory, has to be managed explicitly by the programmer. In return, this
explicit management is often an opportunity for further optimizations and better
performances.

\subsection{GPU-Based Parallel Implementation of the Single-signal Algorithm}
\label {sec:motivation}

In the work presented here we did not produce a parallel implementation of the
single-signal algorithm, but we relied on the results reported in the
literature, instead.

For the parallelization of (single-signal) growing self-organizing network
algorithms, a very common approach is applying the well-known \emph{map-reduce}
pattern, which can be easily parallelized into a two-step method, to the
dominant \emph{Find Winners} phase. In the first step of the \emph{map-reduce}
approach, i.e. the \emph{map} operation, the distance from each unit to the
input signal is computed in parallel. In the second step, i.e. the \emph{reduce}
operation, the set of previously computed distances is iteratively reduced by
comparing subsets in parallel, until the $k$ shortest distances are found. In
passing, Buck \emph{et al.} describe GPU reductions in more detail in the
context of the Brook programming language \cite{buck2004brook}, while Harris
does it in \cite{harrisoptimizing} for the CUDA language. The map-reduce pattern
has been studied in general \cite{liu2011generic} and applied to the search of
$k$ nearest neighbors ($k$-NN) \cite{zhang2012efficient}. More specifically this
approach has been used for the parallelization of the GNG algorithm (see
\cite{garcia2011fast} and \cite{orts2012gpgpu}) and of the Parameter-Less SOM
(see \cite{campbell2005graphics}).

The \emph{map-reduce} approach, however, implies a one-to-one correspondence
between network units and GPU threads in the \emph{map} phase, which becomes
even lower in the \emph{reduce} phase. This fact becomes a substantial
limitation for the parallelization of growing self-organizing networks, which usually start
with a very small number of units and grow progressively during the execution.
As reported (see \cite{garcia2011fast}), unless the network contains at least
500-1000 units, the sequential execution on a CPU can be faster than the
parallel one. To cope with this problem, a hybrid technique has been proposed
(see \cite{garcia2011fast} and \cite{orts2012gpgpu}): switching the execution
from CPU to GPU only when the network is sufficiently large and the latter
hardware is expected to perform better. Nevertheless, even with this hybrid
solution, the maximum level of parallelization that can be attained is bound to
the number of units in the network.

\begin{figure}[t!]
\centering
\includegraphics[width=0.44\columnwidth]{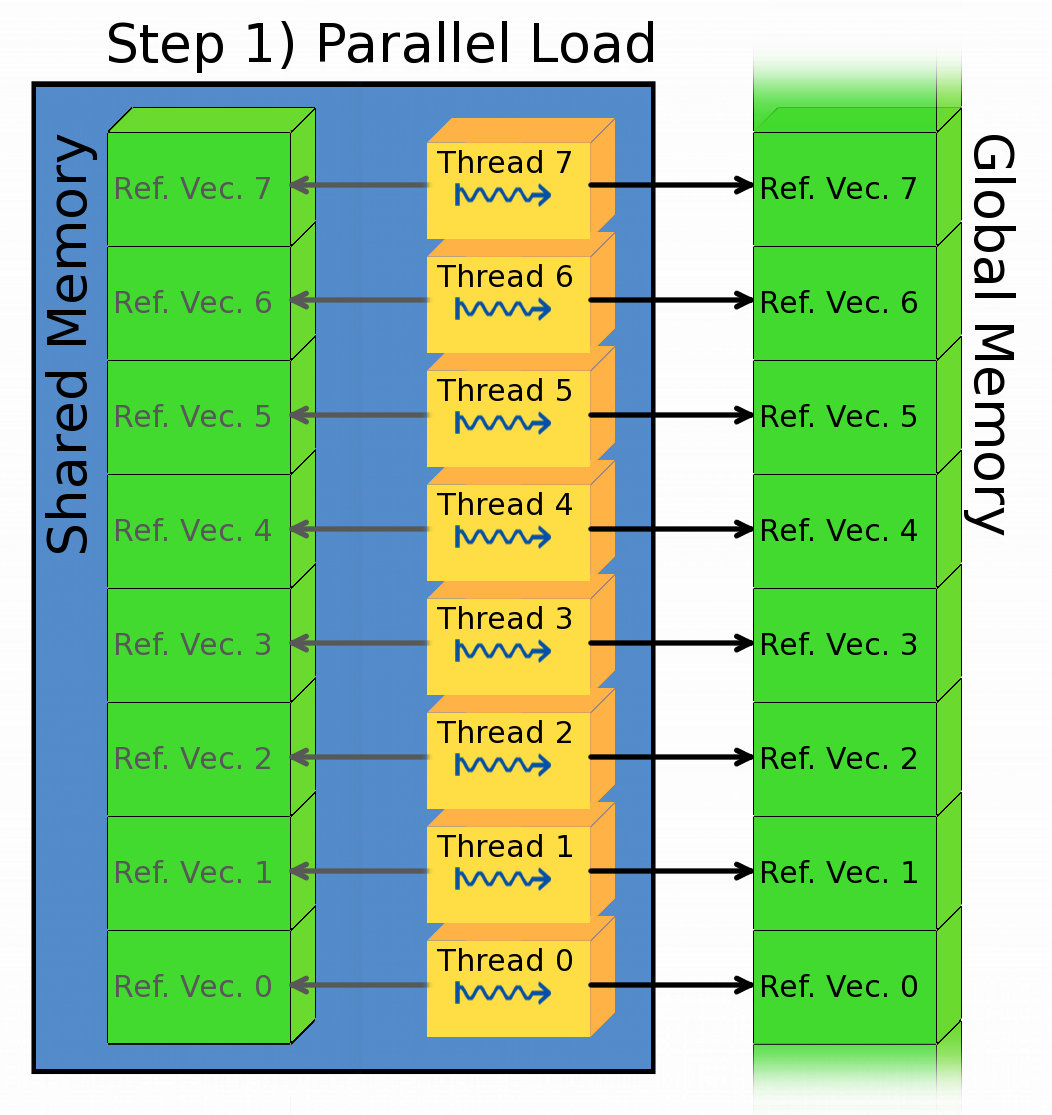}
\includegraphics[width=0.36\columnwidth]{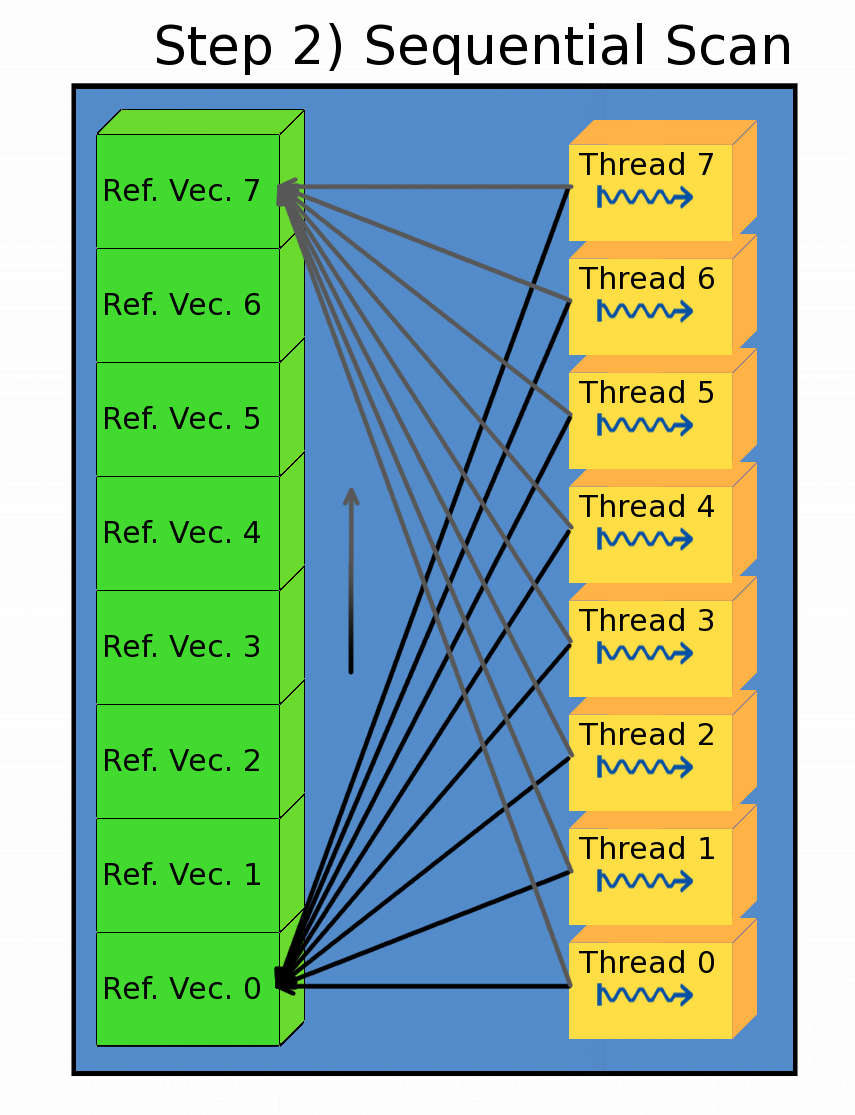}
\caption{The two steps of the Find Winners phase in the parallel
implementation.}
\label{fig:findSteps}
\end{figure}

\subsection{GPU-Based Parallel Implementation of the Multi-signal Variant}
\label{sec:implementation}

For the GPU-based parallel implementation of the algorithm, the main advantage
of the multi-signal variant is that the level of parallelism is bound only by
the number of signals submitted to the network at each iteration. Furthermore
this same level of parallelism can be maintained across entire kernels, since no
reduction takes place. The only limitation of this variant comes from the
collisions that can occur during the \emph{Update} phase, as explained in
section \ref{sec:multi}. Nevertheless, if the parallel implementation focuses on
the dominant \emph{Find Winners} phase, there is in practice no upper limit for
the level of parallelism, beyond that of the hardware.

In the \emph{kernel} that has been realized for the \emph{Find Winners} phase,
each thread is assigned to an input signal. The execution is divided in two
steps (see Fig.\ref{fig:findSteps}): first, all threads in a block load a
contiguous batch of reference vectors in the shared memory with a
\emph{coalesced} access; second, all threads compute the distances from the
reference vectors to the signal through a sequential scan of the shared memory,
in which all threads read the same reference vector in sync. From the point of
view of GPU-based parallelization, this allows harnessing the faster and smaller
\emph{shared memory} in order to accelerate the access to the \emph{global
memory}, i.e. where the whole set of reference vectors is stored.

\section{Experimental Validation}
\label{sec:results}
\subsection{Methods of Comparison}
\label{sec:comparison}

All the experiments described in this section have been performed with the SOAM
algorithm, in four different implementations (see below), applied to the same
tasks of surface reconstruction from point clouds. In each experiment, the point
cloud was taken from a triangular mesh and sampled with uniform probability
distribution $P(\xi)$.

\begin{figure}[t!]
\centering
\subfloat{
\includegraphics[width=0.207\textwidth]{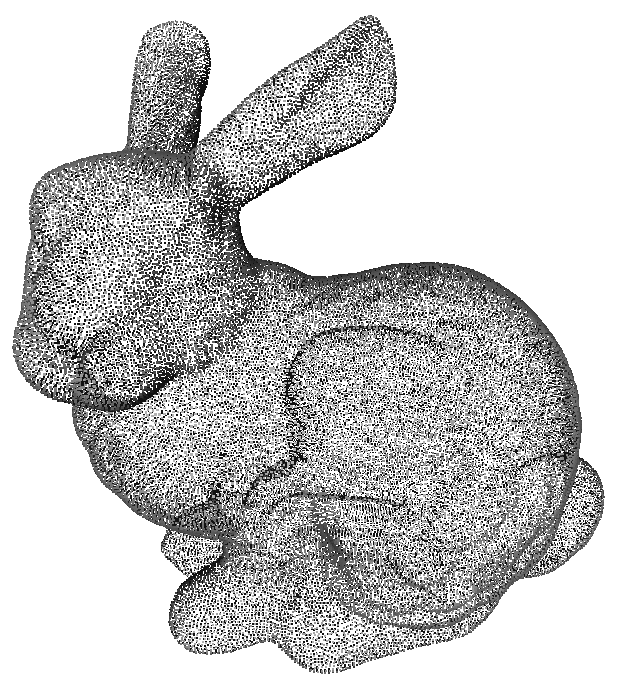}}
\subfloat{
\includegraphics[width=0.126\textwidth]{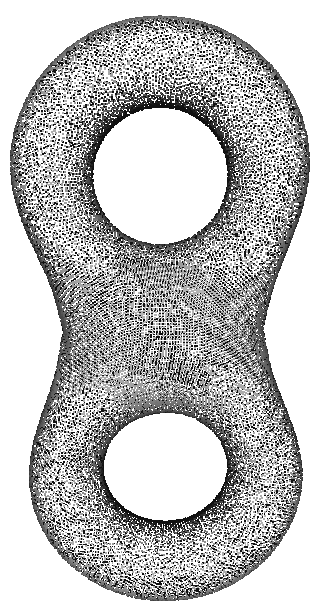}}
\subfloat{
\includegraphics[width=0.297\textwidth]{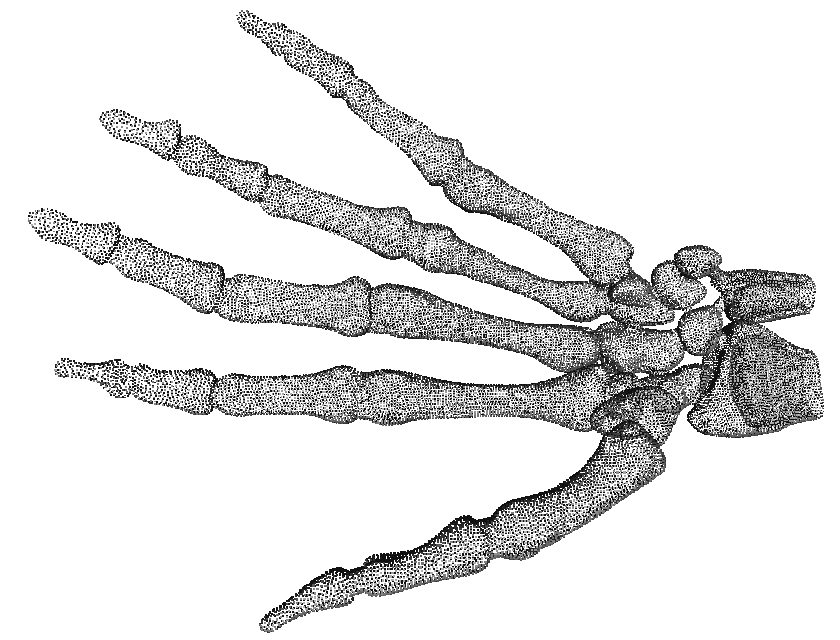}}
\subfloat{
\includegraphics[width=0.243\textwidth]{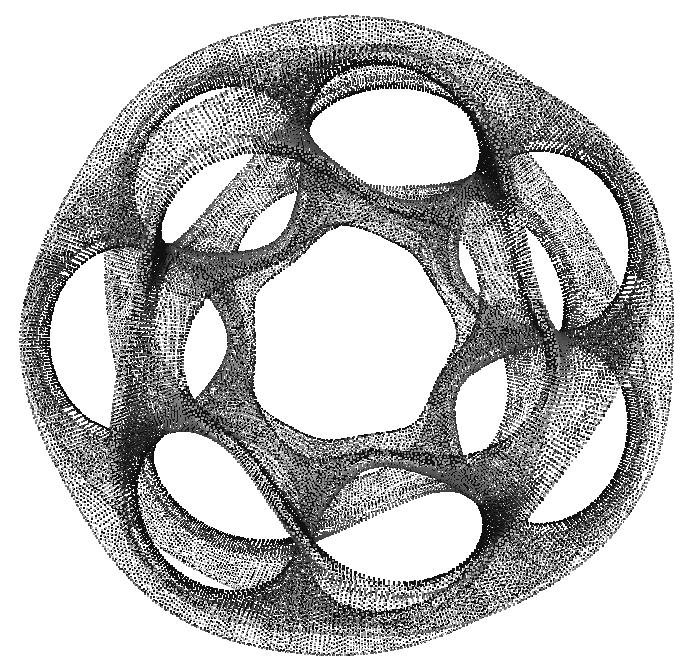}}
\caption{The four point-clouds used in the test phase}
\label{fig:meshes}
\end{figure}

Four different meshes have been used, each having different \emph{topological}
and \emph{geometrical complexity}. More precisely, we consider two measures, one
for each type of complexity: the \emph{genus} of the surface
\cite{Edelsbrunner06}, i.e. the number of holes enclosed by it, and the
\emph{local feature size} (LFS), defined in each point $x$ of the surface as
the minimal distance to the medial axis \cite{amenta1999surface}. In this
perspective, a `simple' mesh has genus zero or very low and high and almost
constant LFS values, while a `complex' mesh has higher genus and LFS values that
vary widely across different areas.
The four meshes used in the experiments are well-known benchmarks for
surface reconstruction (Fig. \ref{fig:meshes}):
\begin{itemize}
  \item \emph{Stanford Bunny}. It has genus $0$ and some
  non-negligible variations in the LFS values that make it non-trivial.
  \item \emph{Eight} (also called \emph{double torus}).
  It has genus 2 and relatively constant LFS values almost everywhere.
  \item \emph{Skeleton Hand}. It has genus 5 and widly variable LFS values, that
  in many areas, e.g. close to the wrist, become considerably low.
  \item \emph{Heptoroid}. It has genus 22, and low and variable LFS values over
  the surface.
\end{itemize}

Obviously, there is no \emph{a priori} guarantee that a parallel algorithm
should be faster than a highly-optimized sequential one. Therefore we chose to
implement also a single-signal variant of the algorithm in which the crucial
\emph{Find Winners} phase is improved through the use of a \emph{hash indexing}
method, similar to that used in molecular dynamics \cite{hockney1992computer}.

The hash index is constructed by defining a grid of cubes of fixed size inside an
axis-parallel bounding box that contains all the input signals in the input
space. The hash index of both signals and reference vectors, which live in the
same space, is obtained from the 3D coordinates. Whenever an input signals is
selected, the search for the reference vectors of the winner and the second
nearest units is first performed on the same cube where the input signal
resides, together with its 26 adjacent cubes. If this search fails, the
exhaustive search is performed instead. Even if this method is
slightly approximate, in that in a few extreme cases it may produce different
results from the exhaustive search, it yields a substantial speed-up, as
will be discussed in section \ref{sec:performances}. In addition, being an hash
method, the maintenance of the index, which is performed in the \emph{Update}
phase, does not affect performances in a significant way.

Four different implementations of the SOAM algorithm have been used for
the experiments:

\begin{itemize}
  \item \emph{Single-signal.} A reference implementation of the single-signal
  SOAM algorithm in C.
  \item \emph{Indexed.} The same single-signal algorithm, but using an hash
  index for the \emph{Find Winners} phase.
  \item \emph{Multi-signal.} A reference implementation in C of the
  multi-signal variant of the algorithm, as described in section
  \ref{sec:multi} and \ref{sec:implementation} but without any actual
  parallelization, in terms of execution.
  \item \emph{GPU-Based.} An implementation in C and NVIDIA C/CUDA of
  the multi-signal variant of the algorithm, with actual hardware
  parallelization.
\end{itemize}

The tests have been performed on a Dell Precision T3400 workstation,
with a NVidia GeForce GT 440, i.e. an entry-level GPU based on the
\emph{Fermi} architecture. The operating system was MS Windows Vista
\emph{Business} SP2 and all the programs have been compiled with MS Visual C++
Express 2010, with the CUDA SDK version 4.0.

\begin{figure}[t]
\centering
\includegraphics[width=0.45\columnwidth]{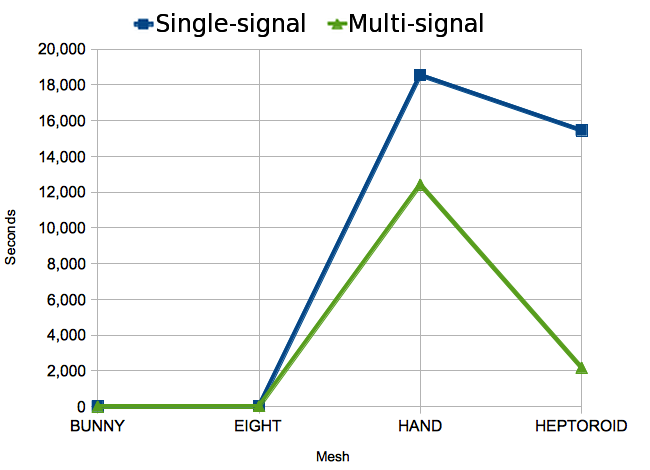}
\caption{Time to convergence of the Single-signal and Multi-signal
implementations.}
\label{fig:totalTimes2}
\end{figure}

All the shared input parameters have been set to the same values for all the
tests for the four different implementations, while implementation-specific
parameters, such as the level of parallelism or the index cube size, have been
tuned specifically for maximum performances. Among the shared input parameters,
only the crucial \emph{insertion threshold} has been tuned for each mesh, for
the reasons described in \cite{piastra2012self}, while every other
parameter value has been kept constant for all the four meshes.

In order to avoid discarding an excessive number of signals in
the \emph{Update} phase, in all parallel implementations the level of
parallelism $m$ at each iteration, i.e. the number of signals processed in the
iteration, is set to the minimum power of two greater than the current number of
units in the network. The maximum level of parallelism has been set to $8192$.

Tables \ref{tab:bunny}, \ref{tab:eight}, \ref{tab:hand}, and
\ref{tab:heptoroid}, at the end of this section, show the numerical results
obtained from the experiments. As it can be seen, for each input mesh, each
different implementation reaches a final configuration which can be either
different or very different, e.g. for the skeleton hand, in terms of number of
units and connections. Note that multi-signal and GPU-based implementations
reach exactly the same final configuration, since they are meant to replicate
the same behavior by design.
As expected, there are substantial differences also for execution times. In the
tables these are measured as \emph{total time to convergence} and
\emph{time per signal}, and the detail figures are reported for each of the
three phases. Time per signal is a measure of the raw performances that can
be obtained with each implementation, while time to convergence is the combined
result of the implementation \emph{and} the different behavior that each
implementation induces, as it will be explained in the next sections.

\begin{figure}[t!]
\centering
\includegraphics[width=0.65\textwidth]{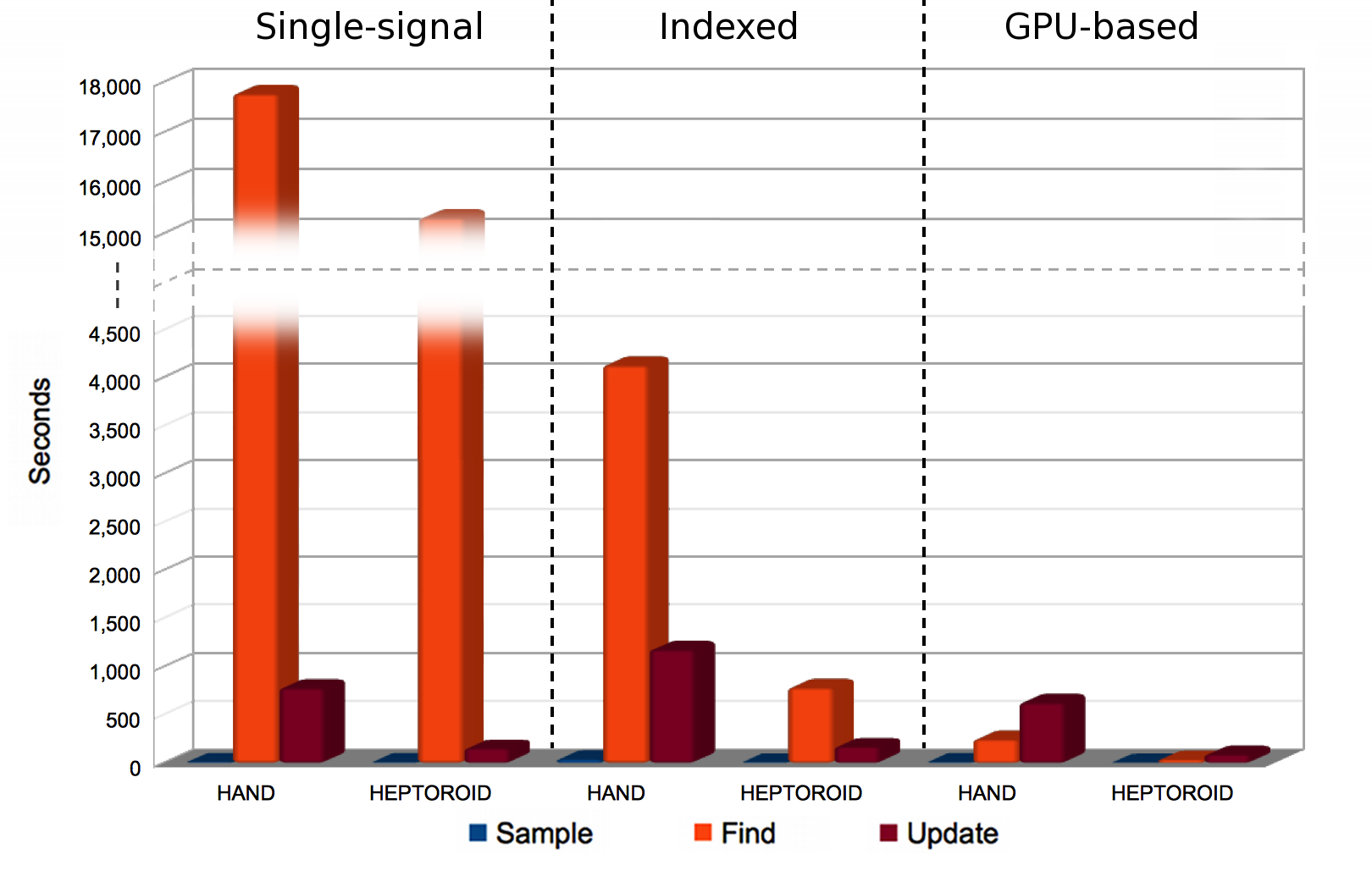}
\caption{Single-phase time to convergence for the two more complex meshes in the
test set.}
\label{fig:phaseTimes}
\end{figure}

\subsection{Behavior of the Multi-signal Algorithm}
\label{sec:behavior}

The first five lines of Tables \ref{tab:bunny}, \ref{tab:eight},
\ref{tab:hand} and \ref{tab:heptoroid}, highlight an aspect that is worth some
further discussion, in particular for the \emph{Single-signal} and the
\emph{Multi-signal} implementations.

Regardless of the hardware parallelizaton, the \emph{Multi-signal} variant
always needs a substantially lower number of input signals than the
\emph{Single-signal} one to converge. This difference becomes even more evident
if the discarded signals are not counted for, approaching a ratio of one to four
as the mesh becomes more complex. The decrease in the number of signals to
convergence is attained despite the growth in the number of units and
connections reached in the final configuration.
Fig.\ref{fig:totalTimes2} compares the times to convergence of the
\emph{Single-signal} and \emph{Multi-signal} implementations. The
performances of \emph{Multi-signal} implementation are always better than
its \emph{Single-signal} counterpart, a difference that becomes more substantial
as the complexity of the mesh increases. Overall, this means that the extra load
due to the increase in the number of both units and connections is outbalanced
by the decrease in the number of signals needed to reach convergence.

In a possible explanation, the multi-signal variant has a better inherently
distributed behavior than the original variant. In fact, in each
iteration of the multi-signal variant, a randomly scattered set of units is
updated `simultaneously', whereas in the single-signal variant only the
winner unit and its direct neighbors are updated. Apparently, the more
distributed update leads to a more effective use of the input signals, yielding
faster convergence. This aspect, however, requires further investigation.

\begin{figure}[t!]
\centering
\subfloat[][Times per signal in the Find Winners phase for the three
implementations.\label{fig:findTimesPerSignal}]{
\includegraphics[width=0.45\textwidth]{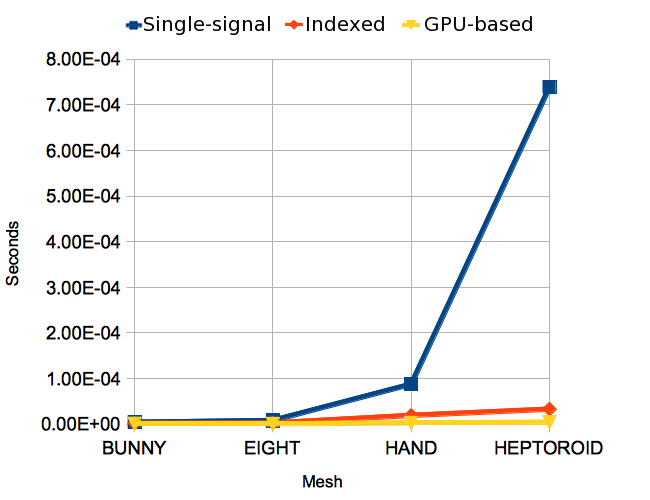}}
\qquad
\subfloat[][Speed-up factors for the Find Winners phase time per signal compared to
the Single-signal implementation. \label{fig:findSpeedupPerSignal}]{
\includegraphics[width=0.45\textwidth]{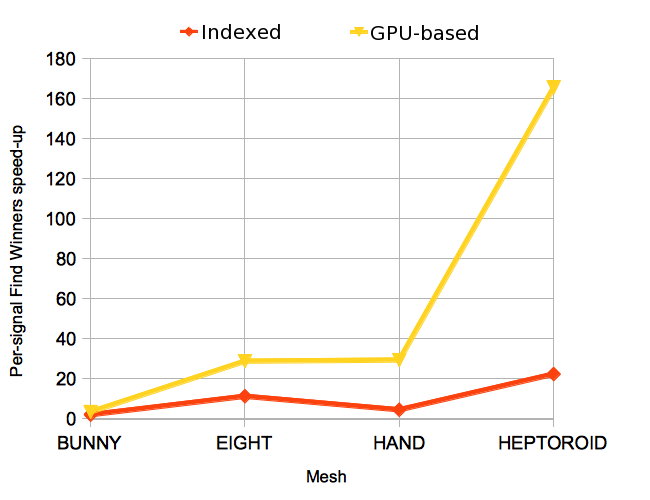}}
\caption{Per-signal performances}
\label{fig:sigPerf}
\end{figure}

\subsection{GPU-Based Implementation Performances}
\label{sec:performances}

Fig.\ref{fig:phaseTimes} shows a summary of the total times to convergence
for the \emph{Single-signal}, \emph{Indexed} and \emph{GPU-based}
implementations respectively, for the two most complex meshes, with detail figure per each
phase. Remarkably, in the \emph{GPU-based} implementation, the \emph{Find Winners}
phase ceases to be dominant, and the \emph{Update} phase becomes the
most time-consuming. This means that in this implementation any further
optimization of the \emph{Find Winners} phase is less relevant unless the
execution of the \emph{Update} phase is sped up in turn.

More in detail, Fig.\ref{fig:findTimesPerSignal} shows the average times per
signal spent in the \emph{Find Winners} phase for each of the three
implementations. Clearly, these times grow as the number of the units in the
network becomes larger. Fig.\ref{fig:findSpeedupPerSignal} compares the speed-up
factors in average time per signal for the \emph{Indexed} and \emph{GPU-based}
implementations with respect to the \emph{Single-signal} one. As expected, these
factors also grow with the number of units in the network, since the hash index
in the \emph{Indexed} implementation becomes more effective, while an higher
level of parallelism can be achieved in the \emph{GPU-based} implementation.
Overall, the speed-up factor for the GPU-based implementation reaches 165x on
the \emph{Heptoroid} mesh.

\begin{figure}[t!]
\subfloat[][Times to convergence for the three
implementations. \label{fig:totalTimes}]{
\includegraphics[width=0.45\textwidth]{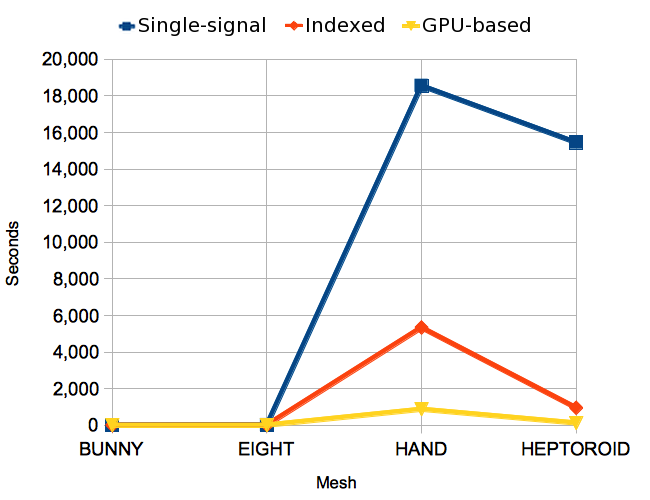}}
\qquad
\subfloat[][Speed-up factors for the time to convergence
compared to the Single-signal implementation.
\label{fig:totalSpeedup}]{
\includegraphics[width=0.45\textwidth]{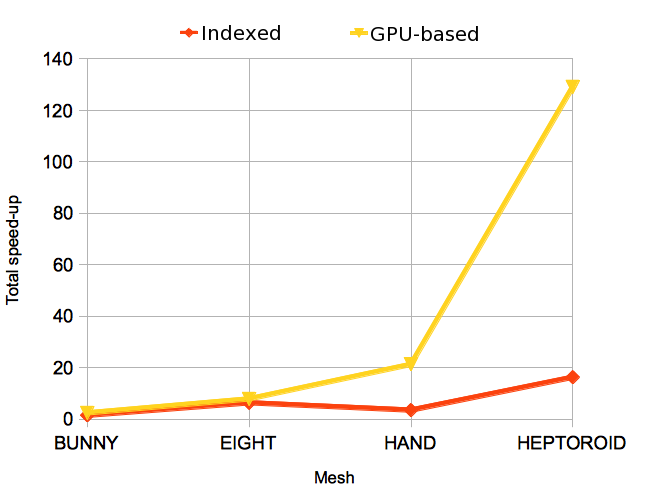}} 
\caption{Global performances}
\label{fig:global}
\end{figure}

Fig.\ref{fig:totalTimes} shows the total times to convergence. These results
show how the performances of the SOAM algorithm depend on the variation in the
LFS values (see section \ref{sec:comparison}): in fact, the skeleton hand always
requires the longest time to convergence, regardless of the implementation.
Fig.\ref{fig:totalSpeedup} shows the speed-up factors for the time to
convergence, for the \emph{Indexed} and \emph{GPU-based} implementations, again
with respect to the \emph{Single-signal} one. These factors grow with the number of
units in the network, and are the combined results of the implementation and of
the behavior induced.

For all input meshes, the total times to convergence for the \emph{GPU-based}
implementation are much lower than the ones for the \emph{Single-signal}
implementation. Speed-ups vary from 2.5x (bunny) to 129x (heptoroid), as the
complexity of the mesh increases and the size of the reconstructed network
grows. In particular, the results obtained with the \emph{Stanford Bunny}, given
in Table \ref{tab:bunny}, show non negligible speed-up factors in both the total
time to convergence (2.5x) and the time per signal (3.5x), despite that the
network contains only 330-347 units at most. This result is particularly
relevant if compared to other GPU-based parallel implementations of growing
self-organizing networks (see for example \cite{garcia2011fast}), for which it
is reported that the GPU execution produces noticeable speed-ups only when the
networks contain no less than 500-1000 units.
The \emph{Indexed} implementation of the algorithm also obtains noticeable
speed-ups on all test cases. Nonetheless, as shown in Fig.\ref{fig:phaseTimes},
the \emph{Find Winners} phase is still dominant, although with \emph{Stanford
bunny} and \emph{Eight} the \emph{Update} times become comparable.

\section{Conclusions and Future Developments}
\label{sec:conclusion}

In this paper we examined the parallelization of growing
self-organizing networks by proposing a multi-signal variant of the original
algorithm adopted, in order to increase its parallel scalability.

In particular, the experiments show that this multi-signal variant adapts
naturally to the GPU architecture in that, besides the advantages deriving from
the careful management of hierarchical memory through perfectly coalesced memory
accesses, it leads to a better usage of the high number of cores by allowing
very high numbers of concurrent threads.

A further interesting, and somehow unexpected, result of the experiments is
that, hardware parallelization apart, the overall behavior of the multi-signal
variant is significantly different from the original, single-signal one. The
multi-signal variant of the algorithm, in fact, seems to better deal with
complex meshes, by requiring a smaller number of signals in order to reach
network convergence. This aspect needs to be further investigated, possibly with
more specific and extensive experiments.

The parallelization described in this paper is limited to the dominant
\emph{Find Winners} phase and, according to the experimental results, can
succesfully make it less time-consuming than the \emph{Update} phase. This means
that future developments of the strategy proposed should aim to the
parallelization of the \emph{Update} phase as well, in order to further improve
on performances. This requires some care however, as the \emph{collisions} among
threads trying to update the data structures involved, must be
managed with care.

\begin{table}
\centering
\small
\caption{Execution time and statistics on the Stanford Bunny data-set.}
\label{tab:bunny}
\centering
\tiny\begin{tabular}{|r||*{4}{r|}}
\hline
\multicolumn{1}{|r||}{\textbf{Algorithm Version}} &
\multicolumn{1}{|c|}{\textbf{Single-signal}}	&
\multicolumn{1}{|c|}{\textbf{Indexed}} &
\multicolumn{1}{|c|}{\textbf{Multi-signal}} &
\multicolumn{1}{|c|}{\textbf{GPU-based}}\\
\hline
\hline
					\multicolumn{5}{|c|}{\bfseries Network Configuration at Convergence}\\
\hline
\multicolumn{1}{|r||}{\bfseries Iterations}	& 620,000 & 616,000	& 1,296 &
1,296\\
\hline
\multicolumn{1}{|r||}{\bfseries Signals} & 620,000 & 616,000 & 580,656 &
580,656 \\
\hline
\multicolumn{1}{|r||}{\bfseries Discarded Signals} & 0 & 0 & 319,054 & 319,054 \\
\hline
\multicolumn{1}{|r||}{\bfseries Units}	& 330 & 332	& 347 & 347 \\
\hline
\multicolumn{1}{|r||}{\bfseries Connections}	& 984  & 990 & 1,035 & 1,035 \\
\hline
\hline
					\multicolumn{5}{|c|}{\textbf{Time to Convergence}}\\
\hline
\multicolumn{1}{|r||}{\bfseries Total Time}	& $4.9530$ & $3.369$ & $3.893$ &
$2.059$\\
\hline
\emph{ Sample}	& $0.460$  & $0.048$  & $0.009$  & $0.016$  \\
\emph{ Find Winners} & $2.610$  & $1.233$  & $2.448$ & $0.699$  \\
\emph{ Update}	& $1.883$  & $2.088$  & $1.436$ & $1.344$  \\
\hline
\hline
					\multicolumn{5}{|c|}{\textbf{Time per Signal}}\\
\hline
\multicolumn{1}{|r||} {\bfseries Time per Signal}	& $7.9887\times 10^{-06}$ &
$5.4692\times 10^{-06}$ & $6.7045\times 10^{-06} $ & $3.5460\times 10^{-06}$\\
\hline

\emph{ Find Winners}	& $4.2097\times 10^{-06}$  & $2.0016\times 10^{-06}$  &
$4.2159\times 10^{-06}$ & $1.2038\times 10^{-06}$
\\
\hline
\end{tabular}
\end{table}

\begin{table}
\centering
\small
\caption{Execution time and statistics on the Eight data-set.}
\label{tab:eight}
\centering
\tiny\begin{tabular}{|r||*{4}{r|}}
\hline
\multicolumn{1}{|r||}{\textbf{Algorithm Version}} &
\multicolumn{1}{|c|}{\textbf{Single-signal}}	&
\multicolumn{1}{|c|}{\textbf{Indexed}} &
\multicolumn{1}{|c|}{\textbf{Multi-signal}} &
\multicolumn{1}{|c|}{\textbf{GPU-based}}\\
\hline
\hline
					\multicolumn{5}{|c|}{\bfseries Network Configuration at Convergence}\\
\hline
\multicolumn{1}{|r||}{\bfseries Iterations}		& 1,100,000 & 1,100,000 & 1,128
& 1,128\\
\hline
\multicolumn{1}{|r||}{\bfseries Signals}			& 1,100,000 & 1,100,000 &
1,100,110 & 1,100,110 \\
\hline
\multicolumn{1}{|r||}{\bfseries Discarded Signals}	& 0		  & 0		& 562,277 &
562,277 \\
\hline
\multicolumn{1}{|r||}{\bfseries Units}			& 656 & 649 & 658 & 658
\\
\hline
\multicolumn{1}{|r||}{\bfseries Connections}				& 1,974 & 1,953 & 1,980 & 1,980
\\
\hline
\hline
					\multicolumn{5}{|c|}{\textbf{Time to Convergence}}\\
\hline
\multicolumn{1}{|r||}{\bfseries Total Time}	& $12.3540$ & $5.5690$ &
$11.6070$ & $3.8690$\\
\hline
\emph{ Sample}	&	 $0.0150$  & $0.0480$  & $0.0620$  & $0.1410$  \\
\emph{ Find Winners}&	 $8.8600$  & $2.8220$  & $8.5060$ & $0.7650$
\\
\emph{ Update}	&	 $3.4790$  & $2.6990$  & $3.0390$ & $2.9630$  \\
\hline
\hline
					\multicolumn{5}{|c|}{\textbf{Time per Signal}}\\
\hline
\multicolumn{1}{|r||}{\bfseries Time per Signal}	& $1.1231\times 10^{-05}$ &
$5.0627\times 10^{-06}$ & $1.0551\times 10^{-05} $ & $3.5169\times 10^{-06}$\\
\hline

\emph{ Find Winners}	& $8.0545\times 10^{-06}$  & $2.5655\times 10^{-06}$  &
$7.7320\times 10^{-06}$ & $6.9539\times 10^{-07}$
\\
\hline
\end{tabular}
\end{table}

\begin{table}
\centering
\small
\caption{Execution time and statistics on the Hand data-set.}
\label{tab:hand}
\centering
\tiny\begin{tabular}{|r||*{4}{r|}}
\hline
\multicolumn{1}{|r||}{\textbf{Algorithm Version}} &
\multicolumn{1}{|c|}{\textbf{Single-signal}}	&
\multicolumn{1}{|c|}{\textbf{Indexed}} &
\multicolumn{1}{|c|}{\textbf{Multi-signal}} &
\multicolumn{1}{|c|}{\textbf{GPU-based}}\\
\hline
\hline
					\multicolumn{5}{|c|}{\bfseries Network Configuration at Convergence}\\
\hline
\multicolumn{1}{|r||}{\bfseries Iterations}		& 202,988,000 & 213,800,000 & 10.264
& 10.264\\
\hline
\multicolumn{1}{|r||}{\bfseries Signals}			& 202,988,000 & 213,800,000 &
81.092.912 & 81.092.912 \\
\hline
\multicolumn{1}{|r||}{\bfseries Discarded Signals}	& 0 & 0 & 33.432.622 &
33.432.622 \\
\hline
\multicolumn{1}{|r||}{\bfseries Units}			& 5,669 & 5,766 & 8.884 & 8.884
\\
\hline
\multicolumn{1}{|r||}{\bfseries Connections}				& 17,037 & 17,328 & 26.688 &
26.688
\\
\hline
\hline
					\multicolumn{5}{|c|}{\textbf{Time to Convergence}}\\
\hline
\multicolumn{1}{|r||}{\bfseries Total Time}	& $18,548.4937$ & $5,337.2451$ &
$12,422.3738$ & $872.0250$\\
\hline
\emph{ Sample}	&	 $9.4050$  & $35.9820$  & $8.6120$  & $8.0480$  \\
\emph{ Find Winners}&	 $17,763.1367$  & $4,127.8511$  & $11,789.8398$ & $241.1750$
\\
\emph{ Update}	&	 $775.9520$  & $1,173.4120$  & $623.9220$ & $622.8020$  \\
\hline
\hline
					\multicolumn{5}{|c|}{\textbf{Time per Signal}}\\
\hline
\multicolumn{1}{|r||}{\bfseries Time per Signal}	& $9.1377\times 10^{-05}$ &
$2.4964\times 10^{-05}$ & $1.5319\times 10^{-04} $ & $1.0753\times 10^{-05}$\\
\hline

\emph{ Find Winners}	& $8.7508\times 10^{-05}$  & $1.9307\times 10^{-05}$  &
$1.4539\times 10^{-04}$ & $2.9741\times 10^{-06}$
\\
\hline
\end{tabular}
\end{table}

\begin{table}
\centering
\small
\caption{Execution time and statistics on the Heptoroid data-set.}
\label{tab:heptoroid}
\centering
\tiny\begin{tabular}{|r||*{4}{r|}}
\hline
\multicolumn{1}{|r||}{\textbf{Algorithm Version}} &
\multicolumn{1}{|c|}{\textbf{Single-signal}}	&
\multicolumn{1}{|c|}{\textbf{Indexed}} &
\multicolumn{1}{|c|}{\textbf{Multi-signal}} &
\multicolumn{1}{|c|}{\textbf{GPU-based}}\\
\hline
\hline
					\multicolumn{5}{|c|}{\bfseries Network Configuration at Convergence}\\
\hline
\multicolumn{1}{|r||}{\bfseries Iterations}		& 20,714,000 & 23,684,000 & 1,244
& 1,244\\
\hline
\multicolumn{1}{|r||}{\bfseries Signals}			& 20,714,000 & 23,684,000 &
7,683,554 & 7,683,554 \\
\hline
\multicolumn{1}{|r||}{\bfseries Discarded Signals}	& 0		  & 0		& 2,262,969 &
2,262,969 \\
\hline
\multicolumn{1}{|r||}{\bfseries Units}			& 14,183 & 13,937 & 15,638 & 15,638
\\
\hline
\multicolumn{1}{|r||}{\bfseries Connections}				& 42,675 & 41,937 & 47,040 &
47,040
\\
\hline
\hline
					\multicolumn{5}{|c|}{\textbf{Time to Convergence}}\\
\hline
\multicolumn{1}{|r||}{\bfseries Total Time}	& $15,449.2950$ & $950.0250$ &
$2,172.8009$ & $119.6530$\\
\hline
\emph{ Sample}	&	 $6.9570$  & $3.4550$  & $0.8010$  & $0.5630$  \\
\emph{ Find Winners}&	 $15,294.3330$  & $780.5370$  & $2,089.6169$ & $34.2640$
\\
\emph{ Update}	&	 $148.0050$  & $166.0330$  & $82.3830$ & $84.8260$  \\
\hline
\hline
					\multicolumn{5}{|c|}{\textbf{Time per Signal}}\\
\hline
\multicolumn{1}{|r||}{\bfseries Time per Signal}	& $7.4584\times 10^{-04}$ &
$4.0113\times 10^{-05}$ & $2.8279\times 10^{-04} $ & $1.5573\times 10^{-05}$
\\
\hline

\emph{ Find Winners}	& $7.3836\times 10^{-04}$  & $3.2956\times 10^{-05}$  &
$2.7196\times 10^{-04}$ & $4.4594\times 10^{-06}$
\\
\hline
\end{tabular}
\end{table}

\bibliographystyle{plain}
\bibliography{document}

\vfill
\end{document}